# Field Angular Momentum Transformation as a Resolution Criterion for Microsphere-assisted microscopy

## Bekirov Arlen*

*Faculty of Physics, Lomonosov Moscow State University, Moscow 119991, Russia*
**bekirovar@my.msu.ru*

## Abstract

This paper presents a sequential analysis of the resolving power of a microparticle as a function of its refractive index and size parameter. We demonstrate that the resolution of a microparticle for incoherent point sources is related to the modal composition of the field, namely the number of excited Mie modes and their amplitudes. Based on this relation, we propose a novel quantitative assessment framework that defines the spatial resolution limit via the ratio of the squared field angular momentum of the source to that of the total particle field (comprising the superposition of the source and scattered fields). The analytical results demonstrate good agreement with both numerical modeling and classical limits.

## 1. Introduction

The optical resolution of a conventional imaging system is fundamentally restricted by the diffraction limit, which is approximately defined as λ/2 (where λ represents the wavelength of light). Overcoming this physical boundary and enhancing spatial resolution remains a critical objective in modern optics and imaging science. One of the most promising avenues to achieve this is microsphere-assisted microscopy [1]. This approach offers significant advantages, as it is a non-invasive technique that can be seamlessly integrated with other advanced modalities, such as confocal, coherent, fluorescence, and dark-field microscopy [2]. This versatility makes microsphere imaging particularly valuable for the high-resolution visualization of delicate biological specimens.

A central and critical question in this field concerns the exact resolving power of the method, its operational limits, and—most importantly—the fundamental factors that restrict its performance. Understanding these limitations is paramount to defining the boundaries of microsphere-assisted imaging. A classical approach to evaluating this resolution involves analyzing how a microsphere images incoherent point sources. These theoretical attempts began almost immediately after the phenomenon was discovered [3]. However, they consistently demonstrated that, except for rare resonant scenarios, the resolving power of a microparticle remains strictly bounded by the classical λ/2 limit. Furthermore, the conditions required to excite these resonances are so specific and stringent that they are nearly impossible to observe or replicate in a practical experimental setup [4]. This theoretical benchmark created a significant paradox in the scientific community, as it directly contradicts numerous successful experimental observations where sub-diffraction resolution was clearly achieved.

To resolve this discrepancy, one must recognize that real-world experiments do not utilize ideal point sources; instead, the illuminating field possesses a degree of coherence, and the target objects have a finite, non-zero size. Following this rationale, a three-dimensional model based on Mie theory was recently proposed [5]. By accounting for realistic illumination conditions and the actual geometry of the visualized specimens, this model demonstrated that partial spatial coherence effects can significantly enhance the resolving power of a microparticle beyond the Abbe limit—without requiring any specific resonant conditions.

Nevertheless, investigating the resolution problem within the classical framework of point-source imaging remains of fundamental importance. According to the Huygens-Fresnel principle, the response of a microparticle to an arbitrary optical field can be rigorously determined as a superposition of its responses to individual point sources [5]. However, a systematic study of this relationship is currently lacking in the literature. More importantly, there are no analytical formulas available to describe this behavior. The present work is specifically dedicated to addressing this gap. Furthermore, experimental data indicate that the highest resolution is achieved when using particles with radii R on the order of a few micrometers. This size range corresponds to a size parameter of $q=2\pi R/\lambda\approx 20$. For this reason, this specific regime will be investigated in comprehensive detail.

# Results and Methods

## 2. The Image Field of a Point Source for a Finite Number of Mie Modes

According to Abbe's theory of image formation, the optical image produced by a thin lens can be understood as a reconstructed Fourier transform of the source field, inherently limited by the truncation of evanescent modes. The objective of this work is to analyze the optical image formed by microparticles, irrespective of the specific optical instrument subsequently used to capture or record the image. To simplify this analysis, it is convenient to utilize an ideal, aberration-free optical system with a magnification of unity. We consider a monochromatic electromagnetic field **E** with a fixed wavelength λ. Under these assumptions, the image field can be mathematically defined using the following expression:

$$\mathbf{E}_{im}\left(x,y,z\right)=\frac{1}{\left(2\pi k\right)^2}\iint\limits_{k_x^2+k_y^2\le(k\,NA)^2}\tilde{\mathbf{E}}^*(k_x,k_y)e^{-i\left(k_x x+k_y y+k_z z\right)}dk_x dk_y\,, \tag{1}$$

where $\tilde{\mathbf{E}}$ represents the two-dimensional Fourier transform of the optical field in the xy-plane, and $k=2\pi/\lambda$ is the free-space wavenumber. The asterisk denotes complex conjugation. The parameter NA is the numerical aperture of the objective lens, classically defined as NA = sin*α*, where *α* is the half-angle of the maximum cone of light that can enter the lens. Equation (1) can be rewritten in real space rather than Fourier space

$$\mathbf{E}_{im}=\frac{k^2}{4\pi}\iint\limits_{\Gamma}(G(\mathbf{n},\nabla)\mathbf{E}^*-\mathbf{E}^*(\mathbf{n},\nabla)G)dS\,, \tag{2}$$

where $G=\exp(ik|\mathbf{r}-\mathbf{r}_0|)/k|\mathbf{r}-\mathbf{r}_0|$ is the Green's function of the scalar wave equation [6], $\nabla=k^{-1}\partial/\partial\mathbf{r}$, and Γ represents the bounding surface that encloses the solid angle cone defined by the numerical aperture NA of the objective. This surface is assumed to be located in the far-field zone, at a distance sufficiently larger than the wavelength from the source. For the limiting case of NA = 1, the surface Γ may be positioned arbitrarily, as long as all sources lie on one side of it.

While Equations (1) and (2) operate within continuous two-dimensional domains in both k-space and real space, respectively, the analysis of microparticle scattering requires a discrete formalization. According to Mie theory [6], the electromagnetic field scattered by a microparticle, $\mathbf{E}^{sca}$, can be rigorously expressed as an expansion over a discrete basis of vector spherical harmonics (VSHs) of the third kind, denoted as **M** and **N** [6]:

$$\mathbf{E}^{sca}=\sum_{l=1}^{\infty}\sum_{m=-l}^{l}a_{lm}\mathbf{N}_{lm}^{(3)}+b_{lm}\mathbf{M}_{lm}^{(3)}\,, \tag{3}$$

When describing the image formation by a microsphere, it is natural to transition from the continuous domains used in Eqs. (1) and (2) to the discrete angular momentum basis ($l$, $m$). To achieve this, we express the image field as an expansion over the same vector spherical harmonics of the first kind, **M** and **N**, which remain finite throughout the entire space. For mathematical convenience, this expansion is formulated using complex-conjugated functions as follows

$$\mathbf{E}_{im} = \sum_{l,m} a_{lm}^{(im)} \mathbf{N}_{lm}^{(1)*} + b_{lm}^{(im)} \mathbf{M}_{lm}^{(1)*} . \tag{4}$$

Due to the linearity of Eqs. (1)–(2), the expansion coefficients of the source field ($a_{lm}$, $b_{lm}$) and the image field ($a_{lm}^{(im)}$, $b_{lm}^{(im)}$) can be related by a matrix equation of the form:

$$\begin{pmatrix} b_{lm}^{(im)} \\ a_{lm}^{(im)} \end{pmatrix} = \mathbf{A}_{im}(NA) \begin{pmatrix} b_{\nu\mu}^{*} \\ a_{\nu\mu}^{*} \end{pmatrix}, \tag{5}$$

The matrix elements of $\mathbf{A}_{im}$ can be obtained directly from Eqs. (1) and (2) by substituting the source field expressed in the form of Eq. (3). For a detailed derivation of the coefficients $\mathbf{A}_{im}$, the reader is referred to the Appendix.

The concept of a "point source" can be physically modeled using an electric dipole [10], whose field can be expressed as:

$$\mathbf{E}^{src} = \mathbf{p}G + \nabla\left(\nabla \cdot \left(\mathbf{p}G\right)\right), \tag{6}$$

where **p** represents the source polarization (dipole moment). In the source plane, the corresponding image fields (for NA = 1) can be derived in a fully analytical form [11]:

$$\mathbf{E}_{im}\left(z = z_0\right) = -i\left( j_0(k\rho)\mathbf{p}^* - \frac{j_1(k\rho)}{k\rho}\mathbf{p}^* + j_2(k\rho)\frac{(\boldsymbol{\rho},\mathbf{p}^*)\boldsymbol{\rho}}{\rho^2} \right) - \frac{J_2(k\rho)}{k\rho}\left( \frac{(\boldsymbol{\rho},\mathbf{p}^*)}{\rho}\mathbf{e}_z + p_z^* \frac{\boldsymbol{\rho}}{\rho} \right), \tag{7}$$

Here $J_n$ are the Bessel functions and $j_n$ are the spherical Bessel functions. For incoherent point sources, the resulting image is formed by the summation of intensities, i.e., $I = I_1 + I_2 \sim |\boldsymbol{E}_{im,1}|^2 + |\boldsymbol{E}_{im,2}|^2$. Usually, the Rayleigh criterion is employed as a quantitative metric for resolution, dictating that the intensity maximum of one point source must coincide with the first intensity minimum of the other. In the classical case, this configuration yields an intensity drop of approximately 20% between the peaks. Following this condition, we assume that two sources are resolvable if the ratio of the total intensity at the midpoint $I_{middle}$ to the peak intensity $I_{max}$ is less than or equal to 0.8, i.e.,

$$I_{middle}/I_{max} \leq 0.8.$$

For vector fields, the spatial resolution can vary depending on the polarization state, a phenomenon previously highlighted in [7]. To eliminate this ambiguity, the image intensity in this study is assumed to be averaged over all polarization states **p** within the *xy*-plane.

The resulting image field distribution is characterized by a finite spatial width on the order of $\lambda/2$. This characteristic scale defines the minimum distance at which two incoherent point sources can be resolved. The physical origin of this limit stems from the fact that the image field, as defined in Eq. (1), is formed exclusively by propagating field components and lacks evanescent spatial harmonics.

As previously noted in the **Introduction**, a microparticle can convert these evanescent waves into propagating ones via scattering. Consequently, to understand the true physical boundaries of this imaging method, one must identify and analyze alternative factors that restrict the resolving power. To address this question, we next examine the source field from Eq. (6) expressed as an expansion over vector spherical harmonics:

$$\mathbf{E}^{src} = \sum_{l=1}^{l_{\max}} \sum_{m=-l}^{l} i \frac{2l+1}{l(l+1)} \frac{(l-|m|)!}{(l+|m|)!} \left[ \left(\mathbf{p}, \mathbf{N}^{(1)}_{l(-m)}(\mathbf{r}_{src})\right) \mathbf{N}^{(3)}_{lm} + \left(\mathbf{p}, \mathbf{M}^{(1)}_{l(-m)}(\mathbf{r}_{src})\right) \mathbf{M}^{(3)}_{lm} \right], \qquad (8)$$

where $\mathbf{r}_{\mathrm{src}}$ is the position vector of the source. In this case the image field can be conveniently evaluated using Eq. (5). Let us examine how the source image evolves as the number of modes in the expansion (8) is truncated. When the number of modes satisfies the condition $l_{\max} > kr_{\mathrm{src}}$, $r_{\mathrm{src}}$ – the distance from the coordinate origin to the source, the image is formed with a diffraction-limited minimal width of $\lambda/2$ (Fig. 1(a)). However, when $l_{\max} < kr_{\mathrm{src}}$, the image undergoes significant additional broadening (Fig. 1(b)). It should be noted that this effect can only be observed at a sufficiently large distance of the source from the origin ($kr_{\mathrm{src}} \gtrsim 10$), since in the limiting case of $kr_{\mathrm{src}}=0$, the number $l_{\max}=1$ and cannot be reduced.

A similar blurring effect can be observed during the formation of a virtual image by the microsphere (Fig. 1(c)). The center of the microsphere is located at the origin, and the source is positioned close to with its surface, i.e., $kr_{\mathrm{src}} \approx kR = q$.

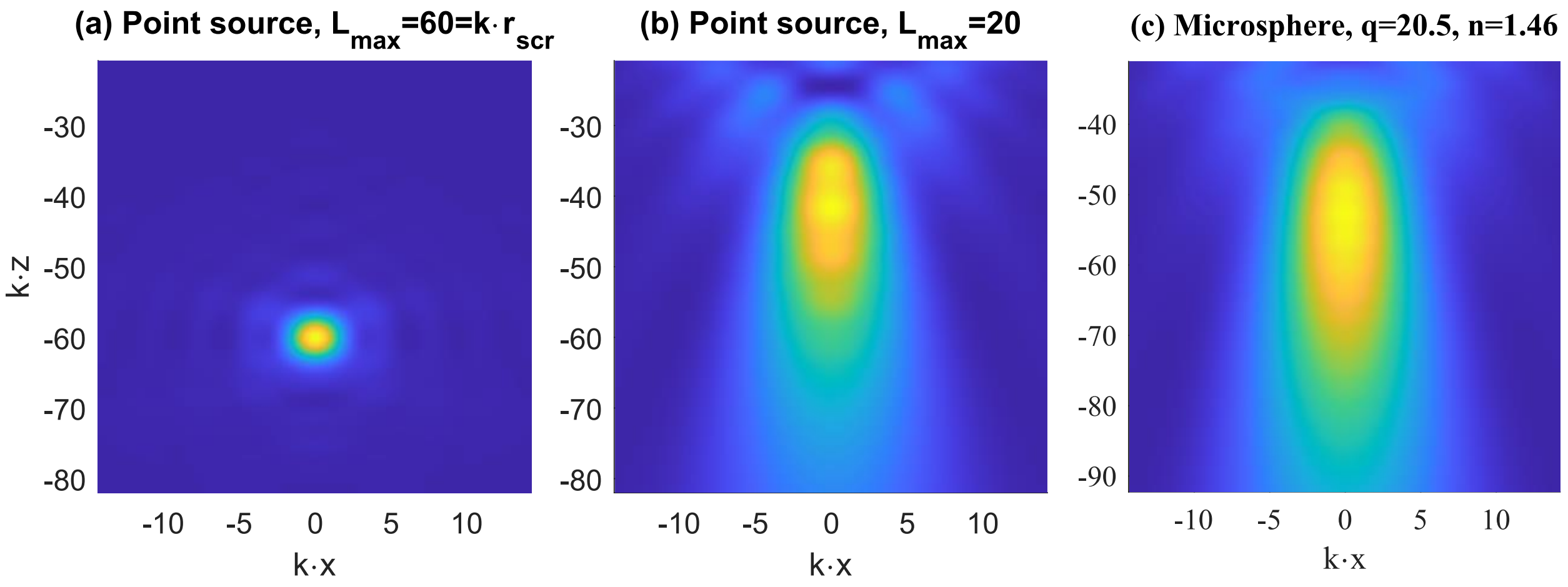

**Fig. 1.** Image fields of a point source located at a distance of $kr_{\mathrm{src}}=60$ under a truncated mode expansion from Eq. (8): (a) $l_{\max}=60$, and (b) $l_{\max}=20$. (c) Image field of a point source positioned directly on the surface of a microsphere with parameter $q = kR = 2\pi R/\lambda = 20.5$, refractive index n=1.46.

By modulating the wavefront of the source field, a transparent microparticle forms an image at a certain distance from its center. For particles with a refractive index of $n$~1.5, a virtual magnified image is produced, with its focus situated further away than the position of the physical specimen. The resulting magnification can be estimated as M~n/(2-n), which yields M~3. Crucially, because the particle is located within the near-field zone of the source, it directly modifies the local field subsequently captured by the microscope objective. This mechanism fundamentally distinguishes microsphere-assisted imaging from conventional post-image magnification, which merely scales the intensity distribution as $I_{im}(Mx, My)$. If we consider the microparticle as an optical system that maps the source positions according to the relation:

$$\mathbf{r}_{src} \rightarrow M\mathbf{r}_{src},$$

where $M$ is the magnification factor, the optical resolution in the object space is then given by $\lambda/(2M)$. This formulation originates from the fact that the minimum resolvable distance in the image space is strictly bounded by the diffraction limit of approximately $\lambda/2$, whereas the magnification $M$ scales the distances between the source images proportionally.

By imposing a constraint on this system due to a finite number of Mie modes, the resulting source field can be described by Eq. (8). Consequently, the resolving power degrades due to the additional blurring in the reconstructed image (Fig. 1(b)). Fig. 2 illustrates the dependence of the system's spatial resolution on the maximum mode number $l_{max}$ at a fixed magnification $M$. The sources are located at a distance $kr_{\mathrm{src}}$=15 from the origin. The behavior presented in Fig. 2 can be accurately approximated by a simple analytical formula. Specifically, in the regime where $l_{max}<Mkr_{\mathrm{src}}$, the resolution asymptotically approaches the expression:

$$\delta \sim \frac{\lambda}{2}\frac{kr_{src}}{l_{max}}. \tag{9}$$

When $l_{max}>Mkr_{\mathrm{src}}$, as noted above, the spatial resolution reaches its limit of $\lambda/(2M)$ and remains constant with any further increase in $l_{max}$. Eq. (9) holds true only for $kr_{\mathrm{src}}\gtrsim 10$, since for smaller values, the image broadening effect resulting from the reduction in the number of modes (Fig.'s 1 (a)-(b)) disappears. The classical Abbe diffraction limit corresponds to the critical mode number $l_{max}=kr_{\mathrm{src}}$, which is determined by the distance from the coordinate origin to the source positions.

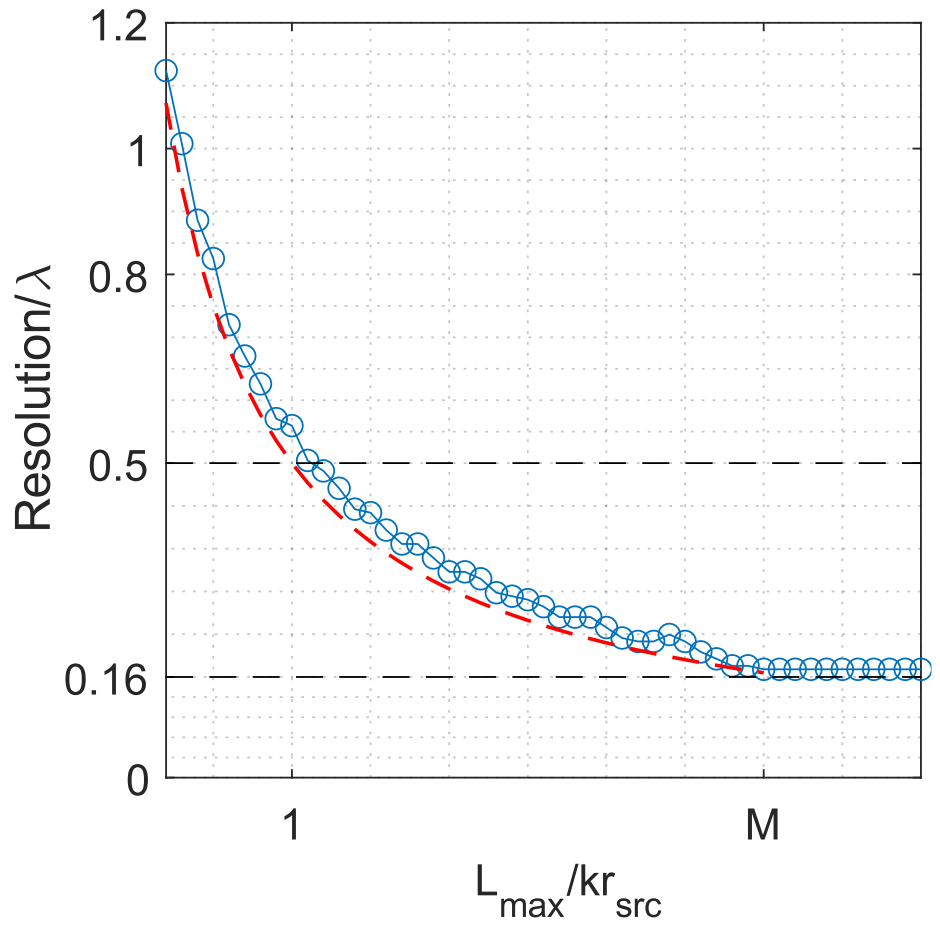

**Fig. 2.** Dependence of the system's resolving power on the number of Mie modes $l_{max}$ at a magnification of M = 3. The red dashed curve corresponds to the asymptotic expression (9). The Abbe limit corresponds to the mode order $l_{max}=kr_{\mathrm{src}}$, which is determined by the distance from the origin to the sources. The maximal resolving power is achieved at $l_{max}=Mkr_{\mathrm{src}}$.

When an image is formed by a microparticle, the distance parameter $kr_{\mathrm{src}}$ cannot be smaller than the size parameter $q=kR$, which represents the limiting case where the source is positioned directly on the microparticle surface. However, the maximum number of modes excited within the microparticle (i.e., those providing a significant contribution to the scattered field) is typically bounded by $l_{max}\sim q$ [8, 9]. According to Eq. (9), this specific condition corresponds to a spatial resolution of $\delta\sim\lambda/2$. Consequently, the image formed by the microsphere undergoes a characteristic additional broadening (Fig. 1(c)), closely resembling the behavior observed in Fig. 1(b). These findings strongly suggest that the finite number of modes excited within a microparticle acts as a fundamental limiting factor that restricts its ultimate resolving power.

## 3. The Role of Focal Plane Selection in High-Resolution Microsphere Imaging.

The exact position of the image plane, $z/R$, is ill-defined for microparticle-assisted imaging, as the spatial intensity distribution exhibits a broad profile along the z-axis (Fig. 1(c)). Furthermore, the profile shape itself heavily depends on the particle size parameter, $q=kR$. According to geometric optics, the image plane position $|z/R|$ is conventionally estimated as $n/(2-n)$. However, when evaluating the maximal resolving power, one must determine an optimal $z/R$ value that corresponds to the specific observation plane where the highest spatial resolution is achieved. This optimal value is not fixed and varies depending on the specific properties of the microparticle. Fig. 3(a) illustrates the resolution dependences calculated for various z/R values within the range of 19 < *q* < 22 for a refractive index of n = 1.46 (corresponding to a glass microparticle). These *q* values are chosen because they approximately correspond to particles with R~2.5 μm, where microsphere-assisted microscopy was first investigated [1]. Numerical calculations demonstrate that, in most cases, the resolving power increases with z/R→-∞. Nevertheless, an excessive increase in the depth of focus gives rise to spurious intensity maxima, causing the images of bright point sources to become surrounded by intense concentric rings. Under such conditions, the reconstructed image fails to accurately reproduce the actual geometry of the studied structure. As the particle size increases, the maximum depth decreases and shifts toward the regime predicted by geometric optics (Fig. 3(b)). It is worth noting that in a real experiment, the practical depth of the observation plane is fundamentally limited by the signal-to-background ratio: while the observable resolution improves with depth, the detected intensity amplitude drops.

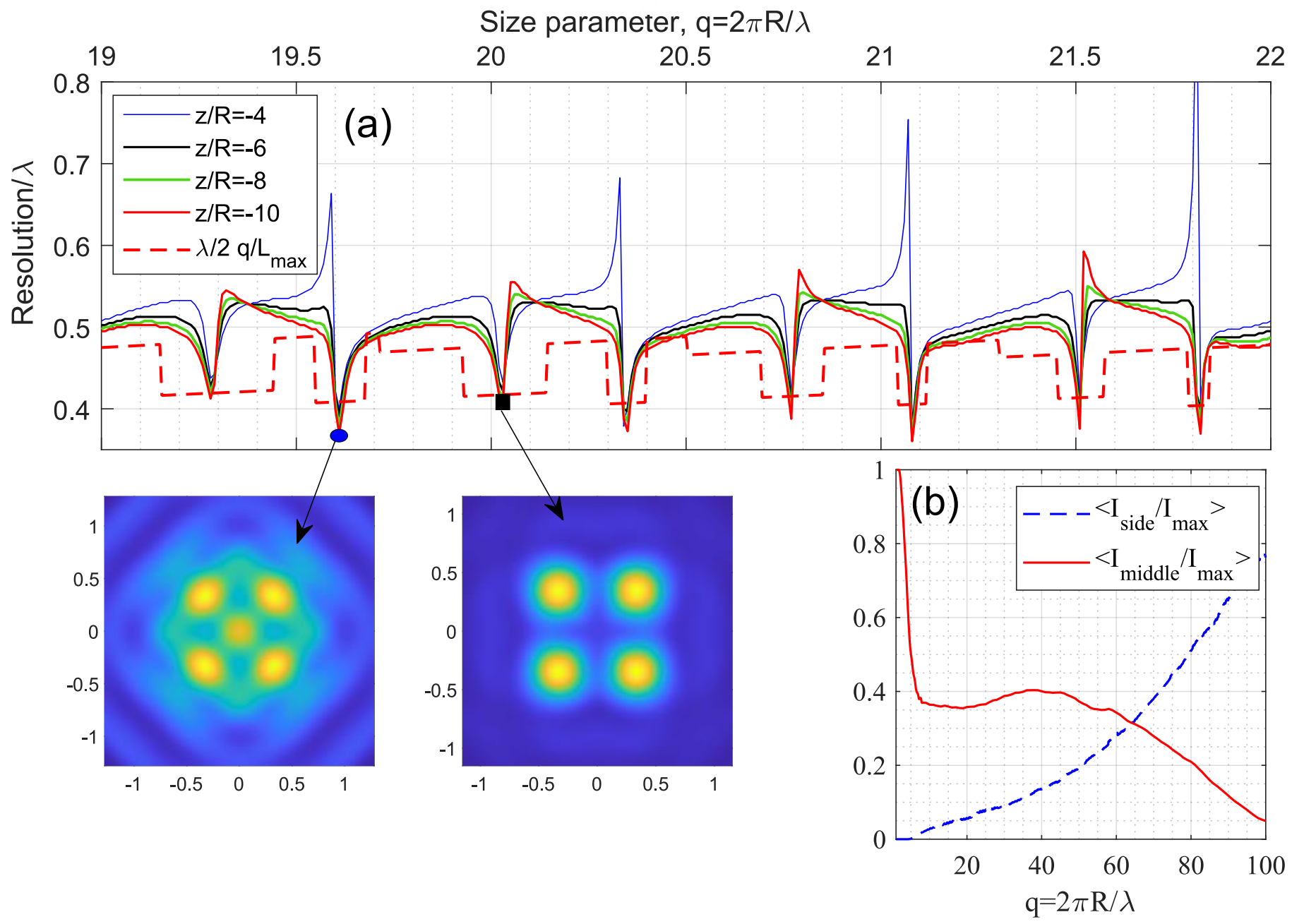

**Fig. 3.** (a) Comparison of the resolving power of a microparticle with a refractive index of *n*=1.46 as a function of the size parameter *q* for various observation planes z/R (for NA = 1). The red dashed curve corresponds to Eq. (10). The blue dot indicates the case where the formally calculated resolution falls below the limit predicted by Eq. (10). When the limit of Eq. (10) is exceeded, the image fails to accurately reproduce the actual structure of the object due to the emergence of secondary side-lobes. The inset shows the reconstructed images of four point sources arranged at the vertices of a square with a diagonal of λ. The dark square marks the point corresponding to the resolution of 0.4λ, where it does not break the limit defined by Eq. (10). (b) Dependence of the averaged ratios $<I_{middle}/I_{max}>_q$ (red curve) and и $<I_{side}/I_{max}>_q$ (blue dashed curve) on the size parameter *q* (for NA = 0.7) of the spherical particle in the z/R = -5 plane for two point sources separated by a distance of 0.6λ, where $I_{side}$ represents the image intensity at the secondary side-lobe.

Fig. 3. (a) Comparison of the resolving power of a microparticle with a refractive index of *n*=1.46 as a function of the size parameter *q* for various observation planes *z/R*. The red dashed curve corresponds to Eq. (10). The blue dot indicates the case where the formally calculated resolution falls below the limit predicted by Eq. (10). This occurs due to the emergence of secondary side-lobes in the image. The inset shows the reconstructed images of four point sources arranged at the vertices of a square with a diagonal of λ. When the limit of Eq. (10) is exceeded, a spurious maximum emerges at the center of the image due to the spatial superposition of side-lobes. The dark square marks the point corresponding to the ultimate resolution of 0.4λ, where it does not break the limit defined by Eq. (10).

## 4. Resolving Power of Microparticles and the Number of Excited Mie Modes

The real resolving power of a microparticle is determined by finding the minimum resolution value across various *z/R* planes. According to Eq. (9), the theoretical resolution limit for a microparticle can be evaluated as:

$$\delta \sim \frac{\lambda}{2}\frac{q}{l_{max}}. \tag{10}$$

Here, the condition $kr_{\text{src}}=q$ corresponds to a source positioned directly on the particle surface, which yields the minimum resolution value. This dependence is plotted in Fig. 3(a) as a red dashed line.

It is important to note that Eq. (10) defines only the potentially achievable resolution. The actual resolving power depends not only on the number of excited modes but also on their complex coefficients, as these modes must interfere in a specific manner to form the final image. The maximum mode order $l_{max}$ is defined by the threshold condition $|a_{l,Mie}|^2+|b_{l,Mie}|^2>0.01$, $a_{l,Mie}$, $b_{l,Mie}$ Mie coefficients [6]. In certain scenarios, particularly under resonance, the numerically calculated resolution shifts below the theoretical limit of Eq. (10). This occurs because the image plane depth *z/R* becomes excessively large. The insets in Fig. 3(a) illustrate the reconstructed image of four point sources arranged at the vertices of a square with a side length of λ. Due to the spatial superposition of secondary side-lobes, a false maximum emerges at the center of the image. Conversely, during resonant excitation—where the resolution does not exceed the limit of Eq. (10)—this artifact is completely absent (see the dark square in Fig. 3(a)), meaning the microparticle delivers the true super-resolution.

## 5. Resolving Power of a Microparticle and the Field Angular Momentum Transformation

The restriction of the microparticle's resolving power by a finite number of excited Mie modes has a fundamental physical interpretation based on the concept of field angular momentum. The indices $l$ and $m$ in the vector spherical harmonics $\mathbf{N}_{lm}$ and $\mathbf{M}_{lm}$ characterize the total angular momentum of the electromagnetic field and its projection, respectively. Consequently, limiting the maximum excited mode order $l_{max}$ is equivalent to bounding the total angular momentum of the scattered field. Therefore, it is appropriate to consider the total angular momentum operator $\boldsymbol{J}$, for whose square operator $J^2$ the functions $\mathbf{N}_{lm}$, $\mathbf{M}_{lm}$ serve as eigenfunctions.

$$J^2\mathbf{N}_{lm},\mathbf{M}_{lm} = l(l+1)\mathbf{N}_{lm},\mathbf{M}_{lm}.$$

Since neither the source field nor the total microparticle field (defined hereafter as the superposition of the source and scattered fields) possess a well-defined total angular momentum, they are represented as linear combinations of the harmonics $\mathbf{N}_{lm}$, $\mathbf{M}_{lm}$. Consequently, one must consider the averaged value of the operator $J^2$:

$$\langle J^2 \rangle = \frac{\int dS\mathbf{E}^* J^2 \mathbf{E}}{\int dS\mathbf{E}^*\mathbf{E}}.$$

Since the optical field **E** is evaluated in the far-field zone, its radial dependence scales as $\sim e^{-kr}/r$ and cancels out when considering the corresponding ratio.

In the scalar approximation, neglecting polarization, the angular momentum of a point source with the field $E^{src}=G(\mathbf{r}, \mathbf{r}_{src})$ can be derived exactly in a closed analytical form:

$$\left\langle J^2_{source}\right\rangle_{scalar} = \frac{\sum_l (2l+1) l(l+1) j_l^2(kr_{src})}{\sum_l (2l+1) j_l^2(kr_{src})} = \frac{2}{3}\left(kr_{src}\right)^2 .$$

When polarization is taken into account, this dependence exhibits a more complex behavior for a source of the form (6). However, for sufficiently large values of $kr_{scr}$, the quadratic growth is preserved with a coefficient approximately of 3/5, i.e.,

$$\left\langle J^2_{source}\right\rangle \approx \frac{3}{5}\left(kr_{src}\right)^2 .$$

Thus, we can conclude that $kr_{scr}\sim\langle J^2\rangle^{1/2}$. The quantity $kr_{scr}$ also characterizes the number of modes within the source, which corresponds to the number of terms in expansion (8) required for the image to be fully formed. This expression can be naturally generalized to the case of a microsphere as:

$$l_{max} \sim \left\langle J^2_{particle}\right\rangle^{1/2} .$$

where $\langle J^2_{particle}\rangle$ is the mean square value of the total angular momentum of the field in the presence of the microparticle (i.e., the superposition of the incident source field and the scattered field). Consequently, Eqs. (8–9) can be rewritten as:

$$\delta \sim \frac{\lambda}{2}\left(\frac{\left\langle J^2_{source}\right\rangle}{\left\langle J^2_{particle}\right\rangle}\right)^{1/2} . \tag{11}$$

Substituting the field expansions from Eq. (3) yields:

$$\delta \sim \frac{\lambda}{2}\left(\frac{\sum_{lm}\xi_{lm}\left(\left|a^{source}_{lm}\right|^2+\left|b^{source}_{lm}\right|^2\right)l(l+1)}{\sum_{lm}\xi_{lm}\left(\left|a^{particle}_{lm}\right|^2+\left|b^{particle}_{lm}\right|^2\right)l(l+1)}\times\frac{\sum_{lm}\xi_{lm}\left(\left|a^{particle}_{lm}\right|^2+\left|b^{particle}_{lm}\right|^2\right)}{\sum_{lm}\xi_{lm}\left(\left|a^{source}_{lm}\right|^2+\left|b^{source}_{lm}\right|^2\right)}\right)^{1/2} , \tag{12}$$

where $\xi_{lm}=4\pi l(l+1)/(2l+1)(l+|m|)!/(l-|m|)!$, while $a^{source}$ and $b^{source}$ represent the expansion coefficients of the source field in the far-field zone (8). Similarly, $a^{particle}=a^{sca}+a^{source}$ and $b^{particle}=b^{sca}+b^{source}$ denote the expansion coefficients of the total field in the outer region of the microparticle. Fig. 4 presents a comparison between the resolving power estimated via Eq. (11) and the numerical calculation. Within the analyzed range, the average resolution value is $\langle\delta\rangle\approx0.45\lambda$, which deviates by only $0.04\lambda$ from the numerically calculated results.

Fig. 4 also presents a comparison with an estimate based on the ratio of the fourth roots of the mean total angular momentum values. This dependence is obtained for a uniform amplitude distribution in the source expansion. Interestingly, while this approach provides better agreement with the numerical calculation in the region of $\delta=0.5\lambda$, Eq. (11) does not reduce to Eqs. (9–10) in this case.

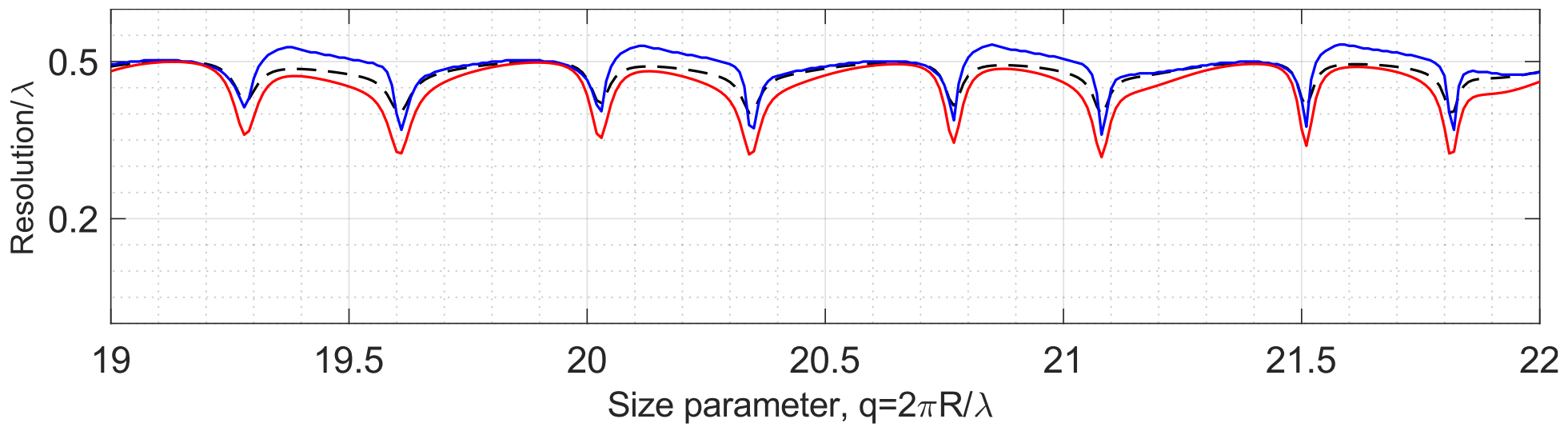

**Fig. 4.** Comparison of the resolving power of a microparticle with a refractive index of $n$=1.46 as a function of the size parameter q. The blue curve represents the numerically calculated resolution values corresponding to the minima of the curves in Fig. 3 for various $z/R$ planes. The red curve shows the resolution estimated via Eq. (11). The black dashed line indicates the dependence obtained when utilizing a power of 1/4 in Eq. (11).

Notably, Eq. (11) correctly describes the limiting case as q → 0. In this regime, the expansion coefficients of the microparticle field approach those of the source field, i.e.,

$$a^{particle} \rightarrow a^{source},\ b^{particle} \rightarrow b^{source}.$$

As a result, the resolution estimate derived from Eq. (11) approaches 0.5λ. Conversely, Eq. (10) formally predicts a zero resolution value in this case (since $l_{max} \geq 1$). Such an estimate lacks physical significance, given that the resolving power in any real optical system remains finite. Furthermore, Eq. (11) successfully captures the resolving power of microparticles with higher refractive indices, such as n > 2. Fig. 5 displays the resolution dependence for a microparticle with a refractive index of n = 6, comparing the prediction of Eq. (11) with the full numerical simulation. In this scenario, the average resolution values are 0.51λ for the numerical calculation and 0.49λ according to Eq. (11). However, the overall agreement is slightly weaker than that observed for particles with $n$=1.46. This discrepancy is primarily attributed to the highly oscillatory behavior of the particle's resolving power and its heightened sensitivity to the selection of the $z/R$ plane.

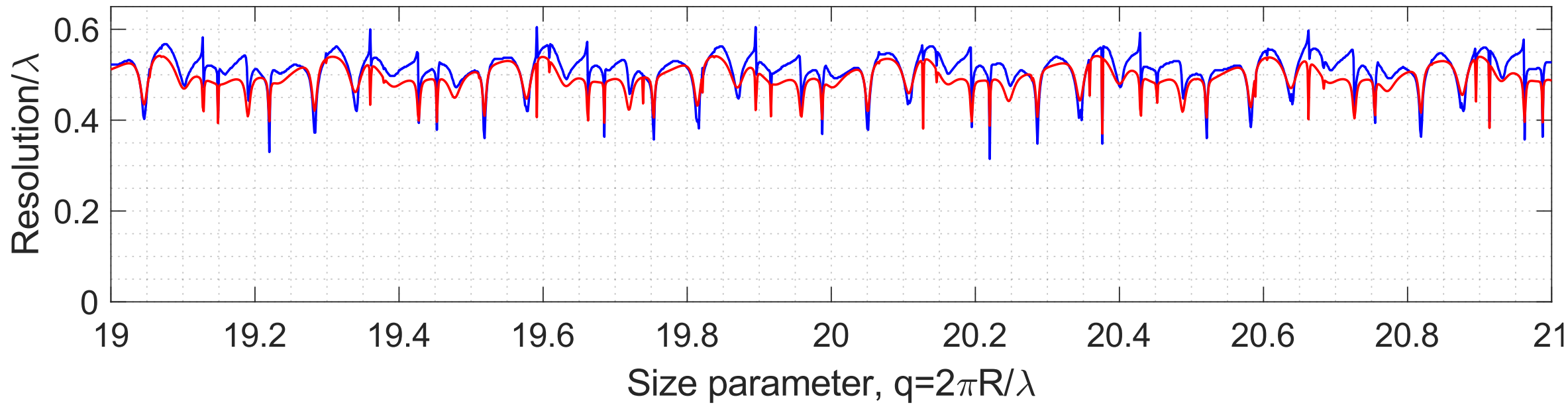

**Fig. 5.** Dependence of the resolving power in the real image on the size parameter $q$ for a particle with $n$=6. The red curve represents the analytical calculation derived from Eq. (11) (with a 1/4 power exponent), while the blue curve shows the full numerical simulation. The image plane was located on the transmission side of the microparticle z>0 and was chosen to correspond to the global resolution minimum.

The generality of the proposed approach is further justified by the fact that the squared total angular momentum operator is intrinsically linked to the expectation value of the field's transverse wavenumber in the far-field zone. Under the Helmholtz equation, the following relation holds true in the far-field region ($r$>>λ):

$$\frac{\langle J^2 \rangle}{r^2} = \langle k_\parallel^2 \rangle$$

Consequently, the ratio of the angular momenta directly characterizes the variation in the transverse components.

*5.1 Transition from Sub-Wavelength Imaging to Geometric Aberration Regime*

Eq. (11) defines the lower bound of the resolving power. The validity of this formula as a lower-bound estimate can only be confirmed if it correctly predicts not only sub-wavelength scenarios $\delta<\lambda/2$ but also regimes where the resolution degrades $\delta>\lambda/2$. Therefore, it is of particular interest to examine cases where Eq. (11) predicts $\delta>\lambda/2$. A prime example is given by particles with large radii ($q>>10$). From the perspective of geometric optics, a spherical particle exhibits significant aberrations, meaning that the resolving power should realistically degrade below the $\lambda/2$ limit in the large-$q$ regime. To clearly visualize this transition, it is convenient to evaluate the averaged resolution values. Fig. 6 compares the resolving power averaged over $\Delta q=1$ for particles with various refractive indices, calculated via Eq. (11). The presented results demonstrate that particles with $n<2$ yield higher resolving power compared to those with $n>2$. Furthermore, a gradual degradation of the resolving power is observed as the size parameter increases.

It should be noted that the applicability of Eq. (11) when $1<q<10$ is not mathematically justified. This limitation arises because its derivation relies on the asymptotic behavior of the angular momentum summation over $l$, which holds true only when a sufficiently large number of modes is involved. This specific regime is indicated in Fig. 6 by a vertical dashed line. Although a microparticle does not form a well-defined image within this range, this region is retained on the plot to ensure a continuous transition between different imaging regimes. The minima observed in Fig. 6 explain why transparent microparticles with a refractive index of $n\sim1.5$ deliver the highest resolving power. According to our findings, this efficiency stems from the microparticle's ability to effectively transform the angular momentum of the optical field. Conversely, the conversion of evanescent waves by the microparticle provides only a potential opportunity to overcome the Abbe limit.

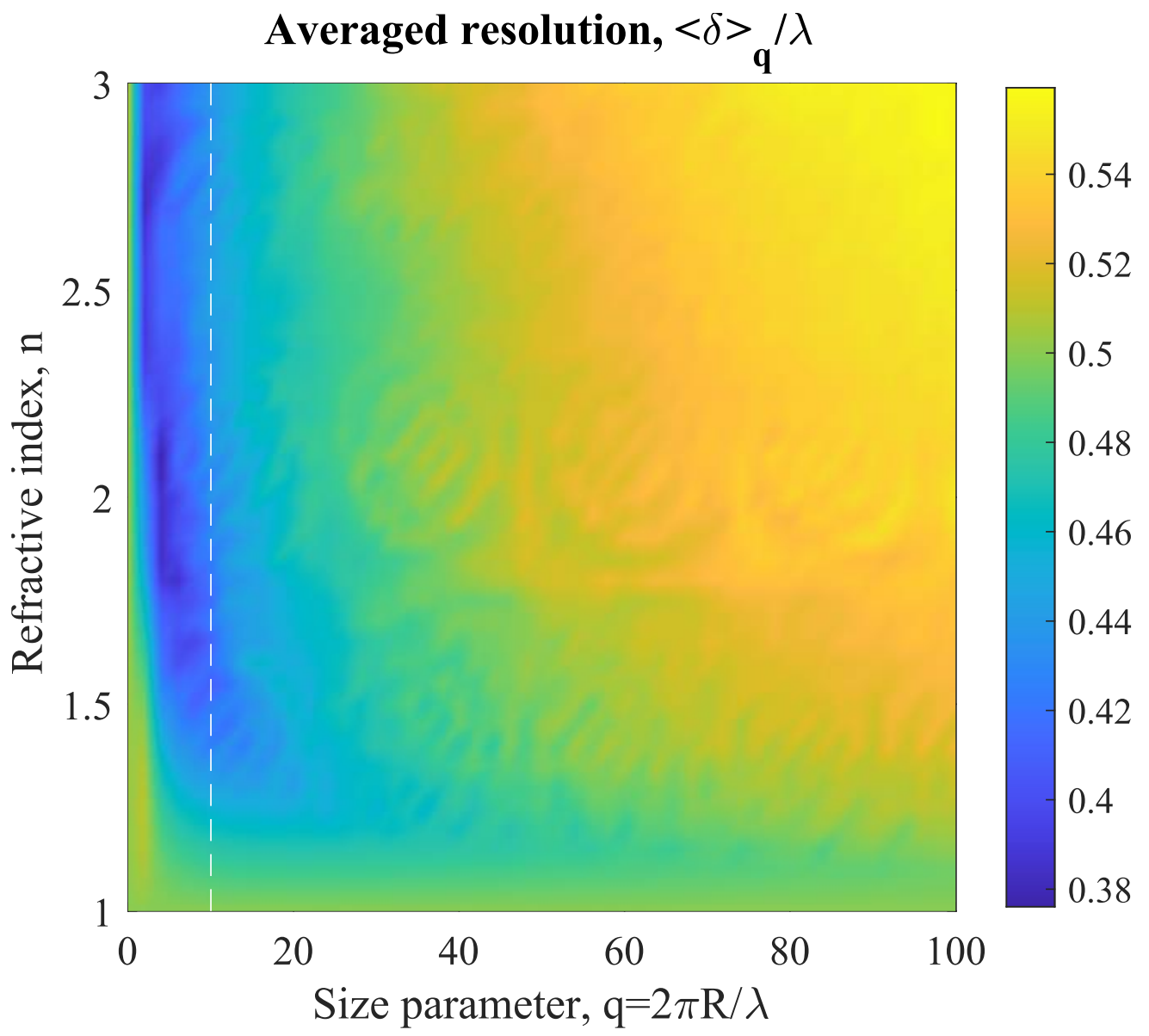

Fig. 6. Dependence of the averaged lower-bound resolution estimate, calculated via Eq. (11), on the size parameter $q$ and the refractive index $n$ of the spherical particle. The color scale corresponds to the normalized resolution estimate $\delta/\lambda$. The white dashed line marks the formal boundary at $q$ = 10, where the particle begins to form a well-defined image.

## 6. Resolving Power Under a Finite Numerical Aperture of the Objective Lens

A fundamental difference between Eq. (11) and the numerical and experimental results is that Eq. (11) does not include the numerical aperture of the microscope objective, NA = sin$\alpha$. Physically, image formation is governed only by the portion of the scattered wavefront that falls within the collection aperture of the microscope. Therefore, to account for a finite numerical aperture, it is natural to replace the averaging over the entire spherical surface by an averaging over its part bounded by the polar angle $\alpha$:

$$\left\langle J^2 \right\rangle_\alpha \sim \frac{\left\langle \mathbf{E} | J^2 | \mathbf{E} \right\rangle_\alpha}{\left\langle \mathbf{E} | \mathbf{E} \right\rangle_\alpha},$$

With this definition, however, the averaged value of the angular-momentum operator$\langle J^2\rangle_\alpha$, generally becomes complex. This is because the operator $J^2$ is self-adjoint only in the function space defined over the complete spherical surface, whereas restricting the integration domain to a spherical segment generally destroys this property. To ensure that the resulting expectation value remains real, we therefore introduce the Hermitian symmetrization of the operator over the restricted integration domain, defined as

$$\left\langle J^2 \right\rangle_\alpha = \frac{1}{2} \frac{\left\langle E | \left(J^2\right)^\dagger + J^2 | E \right\rangle_\alpha}{\left\langle \mathbf{E} | \mathbf{E} \right\rangle_\alpha} \equiv \frac{\mathrm{Re}\left\langle E | J^2 | E \right\rangle_\alpha}{\left\langle \mathbf{E} | \mathbf{E} \right\rangle_\alpha}, \tag{13}$$

where $(J^2)^\dagger$ denotes the Hermitian conjugate operator. By substituting the field expansion in terms of the vector spherical harmonics **M** and **N,** the following expression is obtained in the far-field limit r→∞:

$$\left\langle J^2 \right\rangle_\alpha = \frac{1}{r^2 \left\langle \mathbf{E} | \mathbf{E} \right\rangle_\alpha} \sum_{l,v,m} \frac{(v(v+1)+l(l+1))}{2} \begin{bmatrix} (-i)^{l-v} \left(a_{vm}^* a_{lm} + b_{vm}^* b_{lm}\right) \left\langle \mathbf{n}_{v\mu} | \mathbf{n}_{lm} \right\rangle_\alpha + \\ +(-i)^{l-v+1} \left(a_{vm}^* b_{lm} + b_{vm}^* a_{lm}\right) \left\langle \mathbf{n}_{v\mu} | \mathbf{m}_{lm} \right\rangle_\alpha \end{bmatrix}, \tag{14}$$

where the indices $\mathbf{n}_{lm}$ and $\mathbf{m}_{lm}$ characterize the angular dependencies of the functions $\mathbf{N}_{lm}$ and $\mathbf{M}_{lm}$ in the far-field zone. The overlap integrals of the vector spherical harmonics over the bounded region, which appear in this expression, also determine the elements of the matrix $\mathbf{A}_{im}$ (see Appendix). Direct calculation demonstrates that Equation (14) can be rewritten in the following form:

$$\left\langle J^2 \right\rangle_\alpha = \left\langle \mathbf{A}_{im} J^2 \right\rangle = \frac{\mathrm{Re}\left(\left\langle \mathbf{E} | \mathbf{A}_{im}^*(NA) J^2 | \mathbf{E} \right\rangle\right)}{\left\langle \mathbf{E} | \mathbf{A}_{im}^*(NA) | \mathbf{E} \right\rangle}, \tag{15}$$

where the integration is performed over the entire domain. This equality stems from the fact that the operator $1/2(\mathbf{A}_{im})^*$ acts as a projection operator of the field onto the region of the spherical segment bounded by the numerical aperture NA (see Appendix). Consequently, the fields $J^2\mathbf{E}$ and $1/2\mathbf{A}_{im}^*(\mathrm{NA})J^2\mathbf{E}$ coincide within the region bounded by the NA, whereas outside this region, the field $1/2\mathbf{A}_{im}^*(\mathrm{NA})J^2\mathbf{E}$ is zero. For this reason, the integration domain in Eq (13) can be extended to encompass the entire sphere.

Consequently, by accounting for the finite numerical aperture, Eq. (11) takes the following form:

$$\delta \sim \frac{\lambda}{2}\left(\frac{\left\langle J^2_{source}\right\rangle}{\left\langle \mathbf{A}_{im} J^2_{particle}\right\rangle}\right)^{1/2} . \quad (16)$$

It should be noted that the averaged value of $J^2$ characterizing the source remains unchanged, as it is determined solely by the total number of source modes and is independent of the parameters of the detection system. The influence of the finite numerical aperture is accounted for by the action of the operator $\mathbf{A}_{im}$(NA) on the angular momentum of the particle field.

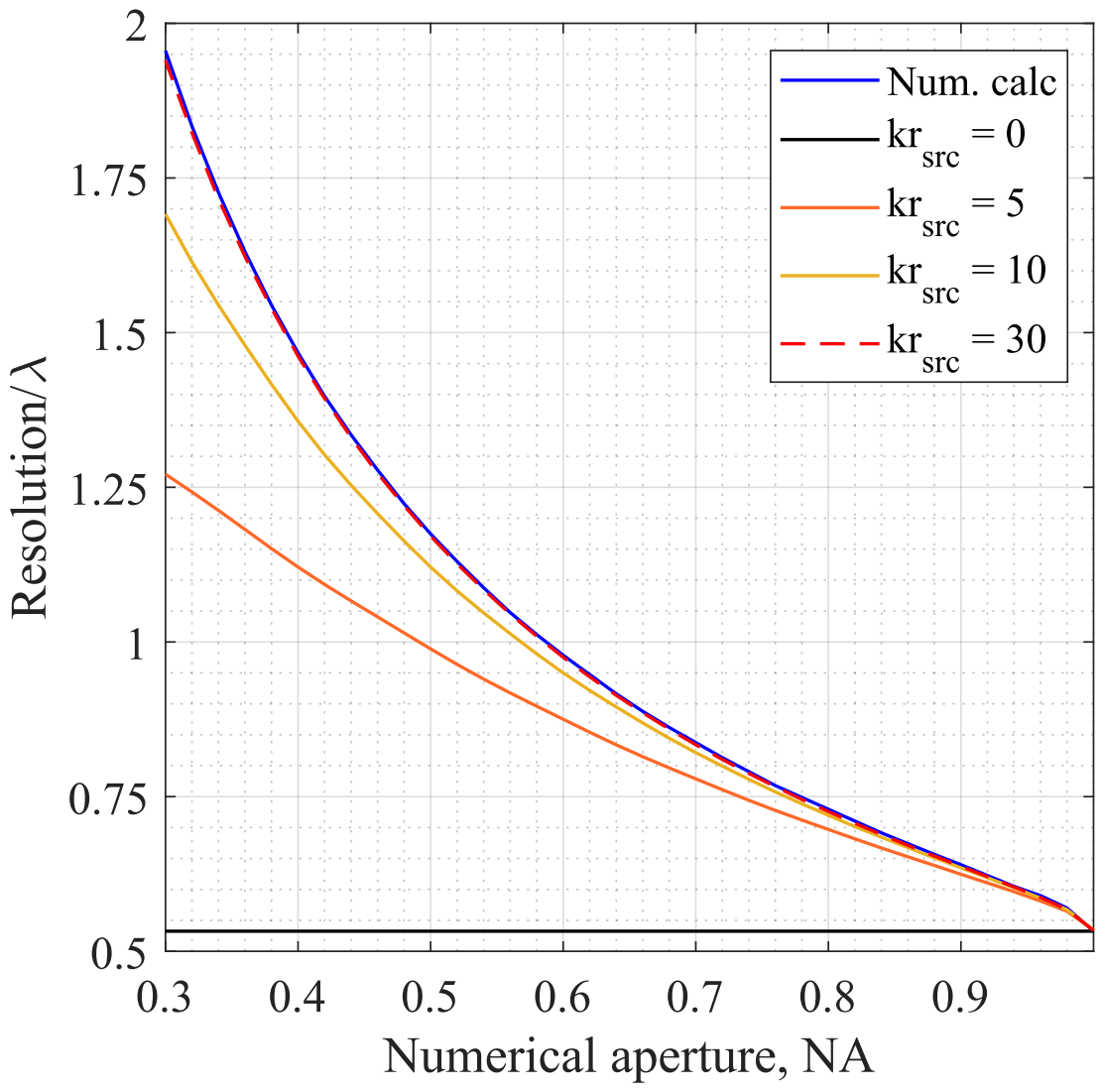

Fig. 7. (a) Comparison of the classical Abbe formula λ/(2sinα) (red curve) with Eq. (16) (blue curve), Eq. (18) (blue dashed curve), Eq. (16) with a power of 1/2 applied to the ratio of moments (black dashed curve). (b) Resolving power dependence for a perfectly electric conducting (PEC) spherical particle, calculated using Eq. (18) (red curve) and Eq. (16) (blue curve).

Fig. 7 presents the dependence of the resolving power, defined by Eq. (16), on the numerical aperture in free space. The numerical calculation was performed by applying the imaging operator $\mathbf{A}_{im}$(NA), followed by the computation of the image field. In this case, the behavior of the numerical curve δ(NA) is independent of the source position $kr_{src}$. However, for the analytical dependence (16), this condition is violated because the concept of resolution limitation due to the truncation of the number of modes breaks down at a small number of modes. As $kr_{src}$ increases, the analytical curve gradually approaches the numerical one. Note that the classical Abbe formula λ/(2NA) in vector theory holds true only for NA≲0.9, whereas Eq. (16) demonstrates good agreement with the numerical calculation across the entire range of NA.

6.1. Effect of local numerical aperture enhancement by a microparticle.

Let us consider the behavior of the resolving power of a microparticle as the numerical aperture (NA) decreases. Experimental results [12] and modeling [5] show that, in contrast to free space, the resolving power of a microsphere is virtually independent of the NA. The fact that the resolution is preserved under a slight reduction in the objective aperture can be confirmed by Eq. (16). Fig. 8 illustrates the resolution dependence for different NA values (1 and 0.8). It can be seen from the plot that the resolution of the microsphere remains nearly unchanged when transitioning from NA=0.8 to NA=1. This phenomenon is commonly referred to as a "local" NA enhancement, since it occurs exclusively for objects located in the immediate vicinity of the microparticle.

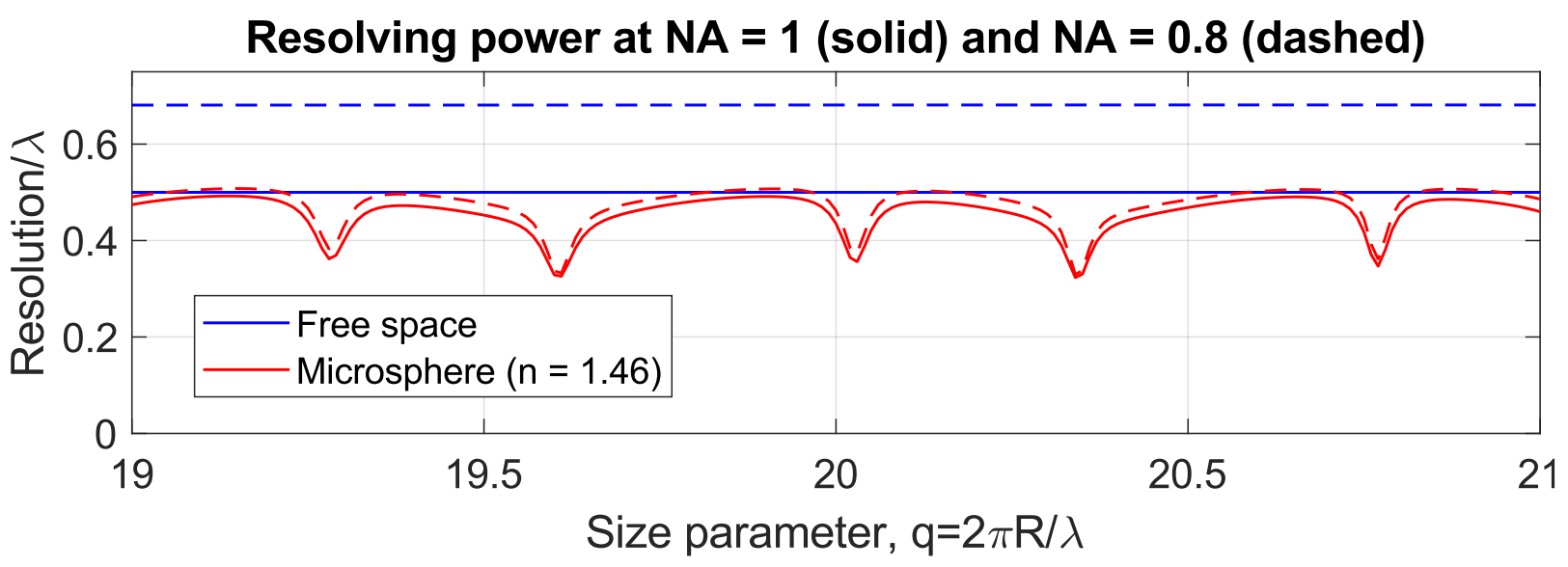

**Fig. 8.** Comparison of the resolving power of a microsphere n = 1.46 and free space at NA=1 (solid curves) and NA=0.8 (dashed curves).

## 7. Imaging Operator Accounting for Magnification and Aberrations

Thus far, image formation has been considered at unit magnification (1:1), independent of the source position. This assumption is convenient for analysing the resolving power and general imaging properties. In practical microscopy, however, objectives typically produce magnified images to match the image scale to the detector resolution or to facilitate observation of the intermediate image through an eyepiece.

According to the Abbe sine condition [13], the transverse magnification is given by the ratio of the numerical apertures at the object and image sides:

$$M_{xy} = \frac{NA_{in}}{NA_{out}},$$

The longitudinal magnification is given by the Herschel condition [13]:

$$M_z = \left( \frac{\sin(\alpha_{in}/2)}{\sin(\alpha_{out}/2)} \right)^2,$$

where $NA_{in}=\sin(\alpha_{in})$ и and $NA_{out}=\sin(\alpha_{out})$, with $\alpha_{in}$ and $\alpha_{out}$ denoting the half-aperture angles on the object and image sides, respectively. For simplicity, the surrounding medium is assumed to have a refractive index of unity.

Physically, magnification arises from the transformation of the field's angular distribution by the objective. In spherical coordinates, this transformation is described by the linear mapping

$$\theta \to \frac{\alpha_{in}}{\alpha_{out}}\theta .$$

This mapping is applied to the far-field angular distribution, making the transformation independent of the arbitrary choice of surfaces used to define the input and output fields.

To incorporate magnification into the present model, we modify the imaging matrix operator $\mathbf{A}_{im}$ by applying the angular-coordinate transformation to its defining components. This yields (see the Appendix):

$$\begin{pmatrix} c_{lm}^{\nu\mu} \\ d_{lm}^{\nu\mu} \end{pmatrix} = \frac{2i}{\xi_{lm}} \begin{pmatrix} i^{\nu-l-1} \left\langle \mathbf{m}_{\nu\mu}(\theta,\varphi) \,|\, \mathbf{m}_{lm}\left( \frac{\alpha_{out}}{\alpha_{in}}\theta, \varphi \right) \right\rangle_{\alpha_{out}} \\ i^{\nu-l} \left\langle \mathbf{m}_{\nu\mu}(\theta,\varphi) \,|\, \mathbf{n}_{lm}\left( \frac{\alpha_{out}}{\alpha_{in}}\theta, \varphi \right) \right\rangle_{\alpha_{out}} \end{pmatrix}.$$

The resulting operator $\mathbf{A}_{im}(NA_{in}, NA_{out})$, accounts for transverse and longitudinal magnification according to the classical Abbe relations. Although analytical evaluation of the resulting integrals over θ is generally difficult, they can be efficiently computed numerically. The form of the vector spherical wave function expansion of the image field, Eq. (4), remains unchanged; magnification is incorporated through modifications of the expansion coefficients $a^{im}$ and $b^{im}$.

The above wavefront transformation describes an idealized case in which points on the input wavefront are mapped uniformly onto the output wavefront according to a prescribed scaling relation. Nonuniform transformations distort the resulting image, giving rise to optical aberrations.

Light passing through a microparticle undergoes scattering and refraction, which modify the wavefront and cause it to deviate from the ideal spherical wave emitted by a point source. These distortions affect the image structure and, consequently, its resolving power. Following the principle of adaptive optics, the aberrations introduced by the microparticle may be partially compensated by appropriately tailoring those of the imaging system, potentially improving the resolution. To account for aberrations, the components of the matrix operator can be calculated as

$$\sim \left\langle \mathbf{m}_{\nu\mu}(\theta,\varphi) \,|\, \mathbf{m}_{lm}\left(f(\theta,\varphi)\right)\right\rangle, \left\langle \mathbf{m}_{\nu\mu}(\theta,\varphi) \,|\, \mathbf{n}_{lm}\left(f(\theta,\varphi)\right)\right\rangle,$$

where $f$ determines the transformation law for the wavefront points at the input and output of the objective. In the general case, this transformation can be non-uniform, thereby modeling aberrations.

As a simple example, consider spherical aberration, which arises from the dependence of the focal position on the ray angle. Partial compensation can be introduced through a nonlinear transformation of the angular coordinate, such as

$$\theta \to \frac{\alpha_{in}}{\sqrt{\alpha_{out}}}\sqrt{\theta}\,. \tag{17}$$

Consider a microparticle with refractive index n=1.46 and size parameter q=20.85. As shown in Fig. 4, the resolving power calculated using the original model differs appreciably from the analytical estimate given by Eq. (11), yielding 0.532λ and 0.484λ, respectively. Incorporating transformation (17) into the imaging operator modifies the image-field distribution and brings the calculated resolving power closer to the analytical estimate, giving 0.487λ (Fig. 9). The calculations use $NA_{in}=NA_{out}=0.95$; choosing values slightly below unity improves the convergence of the numerical integration. Similar improvements are observed for nearby values of $q$.

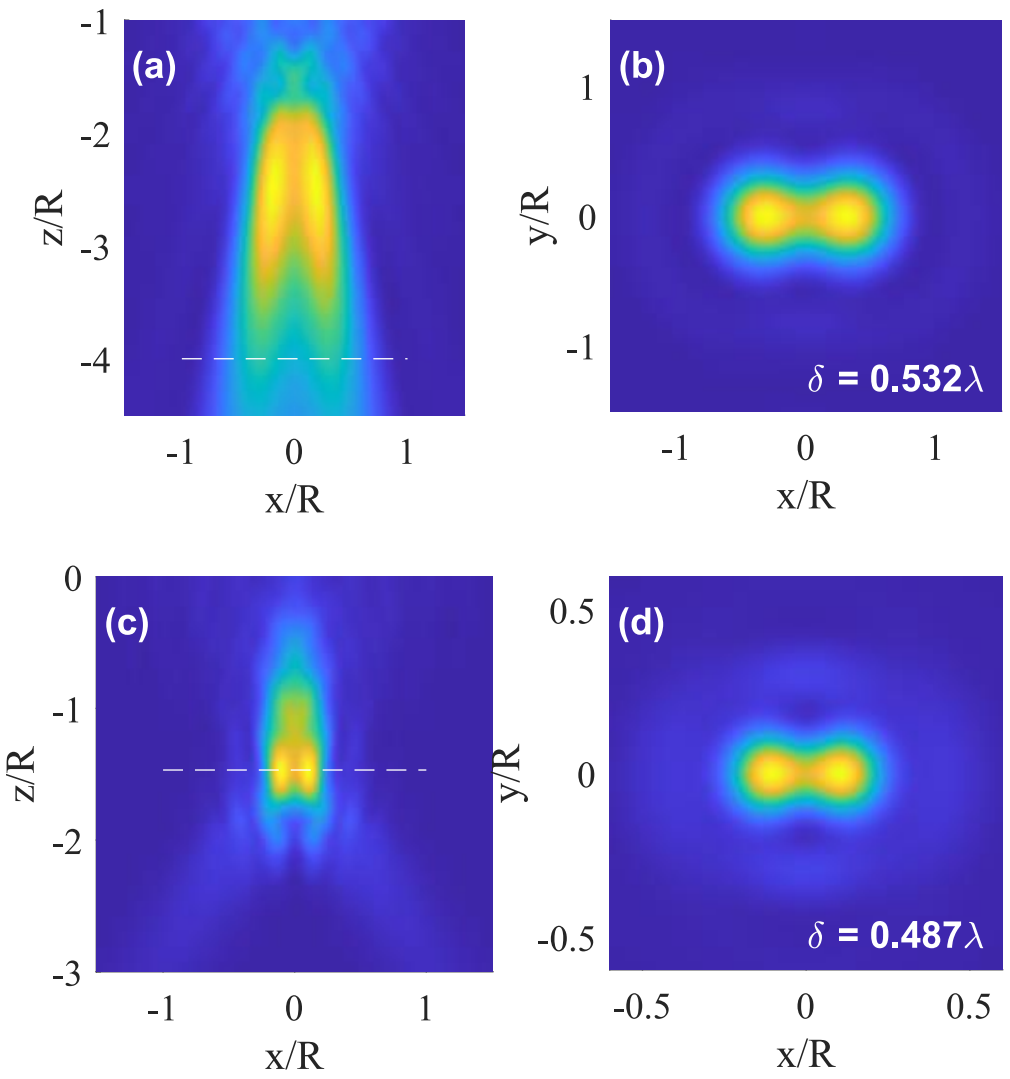

**Fig. 9.** Image-field intensity distribution, $|\mathbf{E}_{im}|^2$, in the *xz* and *xy* planes for two point sources at the resolution limit: (a, b) without aberration correction in the imaging operator $\mathbf{A}_{im}$ (resolving power 0.532$\lambda$); (c, d) with the corrective transformation (17) (resolving power 0.487$\lambda$). Microsphere parameters: $q$=20.85, $n$=1.46. As shown in Fig. 3(a), the resolving power in this case is independent of the image-plane coordinate $z_{im}$.

However, this transformation does not yield the same improvement across the entire particle-size range. For example, at $q$=19, it produces no significant change in resolving power. This indicates that the aberrations introduced by a spherical microparticle depend on its parameters, particularly $n$ and $q$. Thus, the corrective transformation should be tailored to the specific particle parameters to achieve optimal resolution. In general, the optimal transformation may have a more complex functional form than Eq. (17).

# Discussion

The proposed theory should be regarded as a first-order approximation, since it accounts for the transformation of the mean-square angular momentum but does not describe changes in its higher-order moments. Interaction with a microsphere may affect not only the mean-square angular momentum but also the variance of its distribution. One possible explanation is the presence of aberrations introduced by the microsphere, which are absent in free-space propagation. This may explain why Eq. (16) reproduces the numerical results with high accuracy in free space, whereas in the presence of a microsphere, the predicted and simulated trends remain qualitatively consistent but exhibit noticeable quantitative discrepancies. A more general theory should therefore account not only for changes in the mean-square angular momentum but also for changes in its variance and, potentially, other characteristics of its distribution. However, the derivation presented here, which is based on a generalization of Eq. (9), does not provide a framework for such an extension. Developing a more comprehensive approach remains a subject for future research.

Finally, we note that, in free space, the origin of coordinates can be chosen arbitrarily. Accordingly, Eq. (16) can be applied to a source located at an arbitrary position by choosing the coordinate origin appropriately. In the presence of a microparticle, however, the origin must be placed at its center so that the angular momentum characterizes the excitation of Mie modes with respect to the particle center.

# Conclusion

In this work, we have developed a comprehensive theoretical framework to analyze the classical resolving power of microsphere-assisted microscopy. Although the current functional form serves as a first-order approximation with an inherent interpolative nature, it is consistent with rigorous numerical simulations for incoherent point sources, providing a basis for the optimization of super-resolution microsphere imaging systems.

## Appendix

We now turn to the calculation of the components of the operator $\mathbf{A}_{im}$. To obtain the image-field expansion for an arbitrary expansion of the source field, it is sufficient to determine the image field for the functions $\mathbf{M}_{\nu\mu}$. We denote the corresponding field by $\mathbf{E}_{im}(\mathbf{M}_{\nu\mu})$. By linearity, the image field for $\mathbf{N}_{\nu\mu}$ can be obtained by taking the curl of $\mathbf{E}_{im}(\mathbf{M}_{\nu\mu})$. This effectively results in an interchange of the expansion coefficients. We write the expansion of $\mathbf{E}_{im}(\mathbf{M}_{\nu\mu})$ in the following form:

$$\mathbf{E}_{im}(\mathbf{M}_{\nu\mu}) = \sum_{l,m} d_{lm}^{\nu\mu}\mathbf{N}_{lm}^{(1)*} + c_{lm}^{\nu\mu}\mathbf{M}_{lm}^{(1)*} \qquad \text{(A1)}$$

The elements of the sought matrix $\mathbf{A}_{im}$ are the expansion coefficients $d$ and $c$. To calculate these coefficients, we start from the diffraction formula

$$\mathbf{E}_{im}(\mathbf{M}_{\nu\mu}) = -\frac{k^2}{4\pi}\iint_{\Gamma}\left(\left[\mathbf{n},\mathbf{N}_{\nu\mu}^{*}\right]G - i\frac{\mu_0}{n_0}\left(\left[\mathbf{n},\mathbf{N}_{\nu\mu}^{*}\right],\nabla\right)\nabla G + \left[\left[\mathbf{n},\mathbf{M}_{\nu\mu}^{*}\right],\nabla G\right]\right)dS_0,$$

Since this integral does not depend on the choice of the surface $\Gamma$, we let it tend to infinity and choose $\Gamma$ as a segment of an infinite sphere with angular aperture. This angle determines the numerical aperture of the objective: NA=sinα.

Using the expansion of the Green's function G, together with the asymptotic expressions for $\mathbf{M}_{\nu\mu}$ and $\mathbf{N}_{\nu\mu}$ as r→∞,

$$\mathbf{M}_{lm} \to (-i)^{l+1}\,\mathbf{m}_{lm}\,e^{ikr}/kr,$$

$$\mathbf{N}_{lm} \to (-i)^{l}\,\mathbf{n}_{lm}\,e^{ikr}/kr,$$

the expansion coefficients of the field $\mathbf{E}_{im}(\mathbf{M}_{\nu\mu})$ can be written as follows:

$$\begin{pmatrix} c_{lm}^{\nu\mu} \\ d_{lm}^{\nu\mu} \end{pmatrix} = \frac{2i}{\xi_{lm}}\begin{pmatrix} i^{\nu-l-1}\left\langle \mathbf{m}_{\nu\mu} \mid \mathbf{m}_{lm}\right\rangle_{\alpha} \\ i^{\nu-l}\left\langle \mathbf{m}_{\nu\mu} \mid \mathbf{n}_{lm}\right\rangle_{\alpha} \end{pmatrix}. \qquad \text{(A2)}$$

Here, $\xi_{lm}=4\pi l(l+1)/(2l+1)(l+|m|)!/(l-|m|)!$, while the functions **m**, **n** are defined as [1]

$$\mathbf{m}_{lm} = i\pi_{lm}e^{im\varphi}\mathbf{e}_{\theta} - \tau_{lm}e^{im\varphi}\mathbf{e}_{\varphi},$$
$$\mathbf{n}_{lm} = \tau_{lm}e^{im\varphi}\mathbf{e}_{\theta} + i\pi_{lm}e^{im\varphi}\mathbf{e}_{\varphi}.$$

Angle brackets denote integration over the spherical segment:

$$\left\langle \mathbf{f}_1 \mid \mathbf{f}_2 \right\rangle_{\alpha} = \int_0^{\alpha} d\theta \sin\theta \int_0^{2\pi} d\varphi\left(\mathbf{f}_1^{*},\mathbf{f}_2\right).$$

In deriving Eq. (A2), we used the following relations, which can be readily verified by direct substitution:

$$\left\langle \mathbf{m}_{lm} \mid \mathbf{m}_{\nu\mu}\right\rangle = \left\langle \mathbf{m}_{\nu\mu} \mid \mathbf{m}_{lm}\right\rangle = \left\langle \mathbf{n}_{lm} \mid \mathbf{n}_{\nu\mu}\right\rangle = \left\langle \mathbf{n}_{\nu\mu} \mid \mathbf{n}_{lm}\right\rangle,$$

$$\left\langle \mathbf{n}_{lm} \mid \mathbf{m}_{\nu\mu}\right\rangle = \left\langle \mathbf{n}_{\nu\mu} \mid \mathbf{m}_{lm}\right\rangle = -\left\langle \mathbf{m}_{\nu\mu} \mid \mathbf{n}_{lm}\right\rangle = -\left\langle \mathbf{m}_{lm} \mid \mathbf{n}_{\nu\mu}\right\rangle,$$

(the index α is omitted). The integrals over the angular variable $\varphi$ are evaluated straightforwardly. The integrals over $\theta$ are evaluated by integration by parts. Then, using the symmetry with respect to the indices, the difference between the two resulting expressions with different indices is considered. This makes it possible to obtain the following explicit analytical expression:

$$\langle \mathbf{n} | \mathbf{m} \rangle_\alpha = \delta_{m\mu} 2\pi i m \times P_l^m(\cos(x)) P_\nu^m(\cos(x)) \Big|_0^\alpha ,$$

$$\langle \mathbf{m} | \mathbf{m} \rangle_\alpha = \delta_{m\mu} 2\pi \times \left. \frac{\left( l(l+1)\tau_{\nu m}(\cos(x)) P_l^m(\cos(x)) - \nu(\nu+1)\tau_{lm}(\cos(x)) P_\nu^m(\cos(x)) \right) \sin(x)}{l(l+1) - \nu(\nu+1)} \right|_0^\alpha , \text{при } l \neq \nu$$

The last integral for $l=\nu$ is considerably more difficult to evaluate, and obtaining an explicit expression for an arbitrary value of α is challenging. Its evaluation requires a generalization of the Legendre polynomials that allows differentiation with respect to the degree ($l$). The integral can then be evaluated iteratively. The details of the evaluation of this integral are discussed in Refs. [2,3].

At the same time, this integral is readily evaluated for $\alpha=\pi/2$. In this case, due to the symmetry of the functions **m** and **n** with respect to the equator, it is equal to one half of the integral over the full sphere, which is equal to $\xi_{lm}$. This value of $\alpha$ corresponds to the maximum numerical aperture of an objective without immersion, NA=1.

To verify the correctness of our calculations, we compared the image field for source (6) computed via the introduced operator $\mathbf{A}_{im}$ against the analytical formula (7). The proposed approach utilizing $\mathbf{A}_{im}$ demonstrated agreement within machine precision $\sim 10^{-15}$. Furthermore, we performed calculations and compared the images obtained at finite NA<1 values using both (2) and $\mathbf{A}_{im}$, where a high degree of accuracy was also observed.

Since the image field does not allow the source field to be reconstructed uniquely, the following relation follows from its physical meaning:

$$\det \mathbf{A}_{im} = 0 . \tag{A3}$$

The next property of $\mathbf{A}_{im}$ follows from the fact that, if the complex-conjugate field $(\mathbf{E}_{im})^*$ is substituted as the source field **E**, it is mapped back to $\mathbf{E}_{im}$. This is because, by definition, the image field contains no evanescent harmonics, i.e., waves for which $k_x^2+k_y^2>k^2$. This property leads to a specific relation between the columns and rows of the matrix $\mathbf{A}_{im}$.

By definition, the operator $\mathbf{A}_{im}$ acts on the columns of coefficients appearing in the expansion of the source field **E** in spherical functions of the third kind. If the source field is expanded in functions of the first kind, $\mathbf{A}_{im}$ must be modified to $1/2\mathbf{A}_{im}$. We note that the same image field is obtained for a source of the form:

$$\mathbf{E} = \frac{1}{2} \sum_{l,m} a_{lm}^{(im)*} \mathbf{N}_{lm}^{(3)} + b_{lm}^{(im)*} \mathbf{M}_{lm}^{(3)}$$

To prove this statement, $(\mathbf{E}_{im})^* = \mathbf{M}_{\nu\mu}^{(1)}$ should be substituted as the half-sum $\mathbf{M}^{(1)} = ½(\mathbf{M}^{(3)} + \mathbf{M}^{(4)})$ (and analogously for **N**). Then, using the Sommerfeld radiation condition, for the solutions at infinity, the functions of the fourth kind can be discarded because they do not contribute to the image. Physically, this means that waves converging toward the origin do not reach the aperture of the optical system because they propagate in the opposite direction.

Thus, if the operator $\mathbf{A}_{im}$ is applied to both sides of

$$\begin{pmatrix} b_{lm}^{(im)} \\ a_{lm}^{(im)} \end{pmatrix} = \mathbf{A}_{im}(NA)\begin{pmatrix} b_{\nu\mu}^{*} \\ a_{\nu\mu}^{*} \end{pmatrix},$$

with the necessary factor of 1/2 taken into account, the result must be the same expansion of the field $\mathbf{E}_{im}$. Therefore, for any coefficient vector $\mathbf{E}$, one has $\mathbf{A}_{im}E=1/2(\mathbf{A}_{im})^2E$. Hence, $\mathbf{A}_{im}=1/2(\mathbf{A}_{im})^2$. By induction, it can be shown that, for any positive integer $n$, the following relation holds:

$$\mathbf{A}_{im} = \frac{(\mathbf{A}_{im})^n}{2^{n-1}} \tag{A4}$$

It follows from Eq. (A4) that the columns of the matrix are its eigenvectors with eigenvalue 2.

The operator $1/2\mathbf{A}_{im}^*$ acts as a projector onto the spherical segment in the far field bounded by the numerical aperture NA. To demonstrate this, consider a field whose angular dependence is described by the function $\mathbf{M}_{\nu\mu}$, which vanishes outside the spherical segment defined by the polar angle α. Specifically, let

$$\mathbf{M}_{\nu\mu}^{(cut)} \to e^{ikr}/kr\begin{cases} (-i)^{\nu+1}\mathbf{m}_{\nu\mu}, \theta \le \alpha \\ 0, \theta > \alpha \end{cases}.$$

By the completeness of the functions $\mathbf{m}_{lm}$ and $\mathbf{n}_{lm}$, this field can be expanded in terms of these functions. We seek an expansion of the form

$$\mathbf{M}_{\nu\mu}^{(cut)} \to e^{ikr}/kr\sum_{lm} x_{lm}(-i)^l\,\mathbf{n}_{lm} + y_{lm}(-i)^{l+1}\mathbf{m}_{lm},$$

The expansion coefficients are obtained in the standard way:

$$\begin{pmatrix} y_{lm} \\ x_{lm} \end{pmatrix} = \frac{(-i)}{\xi_{lm}}\begin{pmatrix} (-i)^{\nu-l-1}\langle \mathbf{m}_{lm} \,|\, \mathbf{m}_{\nu\mu}\rangle_\alpha \\ (-i)^{\nu-l}\langle \mathbf{n}_{lm} \,|\, \mathbf{m}_{\nu\mu}\rangle_\alpha \end{pmatrix}$$

The resulting expansion coincides exactly with Eq. (A2) for the components of $\mathbf{A}_{im}$, provided that the factor of (1/2) and complex conjugation are included.